\documentclass[%
reprint,
superscriptaddress,
 amsmath,amssymb,
 aps,
prb,
]{revtex4-2}

\usepackage{graphicx}% Include figure files
\usepackage{dcolumn}% Align table columns on decimal point
\usepackage{bm}% bold math
\usepackage{siunitx}
\newcommand{\SRO}{Sr$\mathrm{_2}$RuO$\mathrm{_4}$}
\DeclareSIUnit{\PhiO}{\ensuremath{\Phi_{0}}}
\DeclareSIUnit{\torr}{Torr}
\DeclareSIUnit{\atoms}{atoms}

\begin{document}

\preprint{APS/123-QED}

\title{Local magnetic response of superconducting \SRO{} thin films and rings}%

\author{G. M. Ferguson}
\affiliation{
Department of Physics, Laboratory of Atomic and Solid State Physics,
Cornell University, Ithaca, New York 14853, USA
}
 \altaffiliation[Now at ]{Max Planck Institute for Chemical Physics of Solids}%Lines break automatically or can be forced with \\
\author{Hari P. Nair}
\affiliation{
Department of Materials Science and Engineering, Cornell University, Ithaca, New York 14853, USA
}
\author{Nathaniel J. Schreiber}
\affiliation{
Department of Materials Science and Engineering, Cornell University, Ithaca, New York 14853, USA
}
\author{Ludi Miao}
\affiliation{
Department of Physics, Laboratory of Atomic and Solid State Physics,
Cornell University, Ithaca, New York 14853, USA
}
\affiliation{
Department of Physics, New Mexico State University, Las Cruces, NM 88003
}
\author{Kyle M. Shen}
\affiliation{
Department of Physics, Laboratory of Atomic and Solid State Physics,
Cornell University, Ithaca, New York 14853, USA
}
\affiliation{
 Kavli Institute at Cornell for Nanoscale Science, Ithaca, New York 14853, USA
}

\author{Darrell G. Schlom}
\affiliation{
Department of Materials Science and Engineering, Cornell University, Ithaca, New York 14853, USA
}
\affiliation{
Kavli Institute at Cornell for Nanoscale Science, Ithaca, New York 14853, USA
}
\affiliation{
Leibniz-Institut für Kristallzüchtung, Max-Born-Str. 2, 12489 Berlin, Germany
}
\author{Katja C. Nowack}%
 \email{kcn34@cornell.edu}

 \affiliation{
Department of Physics, Laboratory of Atomic and Solid State Physics,
Cornell University, Ithaca, New York 14853, USA
}
\affiliation{
 Kavli Institute at Cornell for Nanoscale Science, Ithaca, New York 14853, USA
}


\begin{abstract}
We conduct local magnetic measurements on superconducting thin-film samples of \SRO{} using scanning Superconducting Quantum Interference Device (SQUID) susceptometry. From the diamagnetic response, we extract the magnetic penetration depth, $\lambda$, which exhibits a quadratic temperature dependence at low temperatures. Although a quadratic dependence in high-purity bulk samples has been attributed to non-local electrodynamics, our analysis suggests that in our thin-film samples the presence of scattering is the origin of the quadratic dependence. While we observe micron-scale variations in the diamagnetic response and superconducting transition temperature, the form of the temperature dependence of $\lambda$ is independent of position. Finally, we characterize flux trapping in superconducting rings lithographically fabricated from the thin films, paving the way to systematic device-based tests of the superconducting order parameter in \SRO{}.
\end{abstract}

\maketitle

% To-do and questions (numbering the items so it's easier to discuss them)

%9) Fix/look at references to Appendix and Supplementary Information. Are they correct/consistent etc.

% Introduction to SRO214
Since the discovery of superconductivity in \SRO{} \cite{maeno1994superconductivity}, a substantial research effort has been dedicated to the determination of the superconducting order parameter \cite{ishida1998spin, luke1998time, nelson2004odd, xia2006high, kidwingira2006dynamical}. Although early experiments were interpreted in favor of an odd-parity, time reversal symmetry breaking order parameter, recent experiments have called this picture into question \cite{hicks2014strong, steppke2017strong, pustogow2019constraints, ishida2020reduction, li2021high}. High-quality superconducting thin films of \SRO{} have recently been grown by molecular beam epitaxy \cite{nair2018demystifying}, raising the prospect of device-based tests of the order parameter symmetry \cite{tsuei2000pairing}. Given the extreme sensitivity of the superconducting state in \SRO{} to disorder \cite{mackenzie1998extremely}, a direct comparison between the superconducting properties of thin films and bulk single crystals is desirable.

% Maybe expand? Thin films make strain engineering possible 

% Introduction to the magnetic penetration depth
The dependence on temperature, $T$, of the magnetic penetration depth, $\lambda$, contains information about the gap structure of superconductors. A nodal superconducting gap gives rise to a power-law dependence of $\lambda(T)$ at low temperature, while a fully-gapped superconductor exhibits an exponential temperature dependence. In \SRO{}, the observation of $\lambda(T) \sim T^{2}$ in high-quality bulk single crystals provided early evidence for the presence of nodes in the superconducting gap function \cite{bonalde2000temperature}. 

% Introduction to thin films and the sample that we measure
Here, we study the superconducting penetration depth in a \SI{30}{\nano\meter} thick \SRO{} thin film grown by molecular-beam expitaxy. The substrate is $(\mathrm{LaAlO_3})_\mathrm{0.29}-(\mathrm{SrAl}_{0.5}\mathrm{Ta}_{0.5}\mathrm{O}_3)_{0.71}$ (LSAT) with the c-axis oriented perpendicular to the substrate surface. The LSAT substrate preserves the tetragonal symmetry of \SRO{} and induces a small tensile strain of approximately $0.045\%$ when the sample is cooled to low temperatures. Details of the growth of these films are described in Ref.~\cite{nair2018demystifying} and in Appendix C.Micrometer-scale rings were fabricated using photolithography and argon ion milling from a second thin film with thickness \SI{30}{\nano\meter} grown on LSAT (see Appendix D for details on the fabrication).

\begin{figure}
    \centering
    \includegraphics[width=0.45\textwidth]{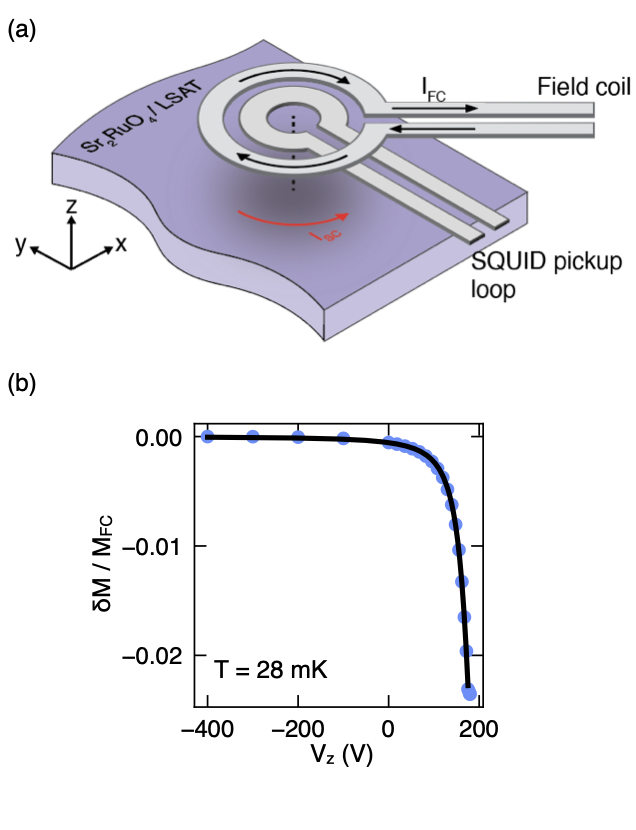}
    \caption{(a) Schematic of the SQUID pickup loop and field coil pair close to the surface of an \SRO{} thin film grown on an LSAT substrate. A current $I_{FC}$ applied to the field coil with an inner diameter of \SI{8}{\micro\meter} induces a screening current $I_{SC}$ in the superconducting \SRO{}. The sample response to $I_{FC}$ is detected via the SQUID pickup loop, with an inner diameter of \SI{1.5}{\micro\meter}. (b) Change in mutual inductance $\delta M$ between the SQUID and field coil as the voltage $V_z$ applied to the z-positioner is increased, bringing the susceptometer towards the sample surface. The solid line is a fit to Eq.~\ref{eq:eq1}}.
    \label{fig:fig1}
\end{figure}

% Introduction to our experiment
Our experimental approach is illustrated schematically in Fig. \ref{fig:fig1}a. We use a scanning SQUID susceptometer \cite{gardner2001scanning,huber2008gradiometric} to detect the diamagnetic response of the superconducting film to a local magnetic field produced by a current $I_{FC}$ sourced through an integrated field coil with radius $a$. In the superconducting state, the \SRO{} film generates supercurrents that screen the magnetic field produced by the field coil, effectively reducing the mutual inductance $M$ between the field coil and the SQUID pickup loop.

 In Fig.~\ref{fig:fig1}b we show the change in the measured mutual inductance $\delta M / M_{FC}$ at a fixed temperature, as the susceptometer approaches the sample surface. Here, we set $\delta M = 0$ when the SQUID is far from the sample, and $M_{FC}$ is the bare mutual inductance between the SQUID pickup loop and field coil, which we measured to be \SI{330}{\PhiO \per\ampere} far from the sample surface. With these choices, $\delta M / M_{FC} = -1$ would correspond to a complete screening of the magnetic field in the plane of the pickup loop. The signals that we observe correspond to a fraction of this value as expected for a thin film. In the Pearl limit $\lambda >> d$ with $d$ the film thickness and neglecting the finite thickness and detailed geometry of the pickup loop and field coil, Kogan and Kirtley \textit{et al.} derived a model for the dependence of $\delta M / M_{FC}$ on the height $z$ above the sample:
\begin{equation}
    \frac{\delta M(z, T)}{M_{FC}} = -\frac{a d}{2 \lambda^2(T)} \left[ 1 - \frac{2 z}{\sqrt{a^2 + 4z^2}}\right].
    \label{eq:eq1}
\end{equation}

In Fig.~\ref{fig:fig1}b we include a fit to Eq.~\ref{eq:eq1} of the height dependence of $\delta M / M_{FC}$. We constrain the fit parameters to specific values: $a = \SI{4}{\micro\meter}$, $d = \SI{30}{\nano\meter}$ and $z_0 = \SI{4}{\micro\meter}$, where $z_0$ is the height of the susceptometer above the sample surface when the corner of the SQUID chip first makes mechanical contact with the sample. The estimated value of $z_0$ is based on our alignment and the SQUID geometry. The fitting procedure yields $\lambda = \SI{350}{\nano\meter}$ and $\alpha = \SI{120}{\nano \meter \per\volt}$. Here, $\alpha$ characterizes the change in SQUID height per volt applied to the piezo positioner. These values agree reasonably well with previous measurements of $\alpha$ for our scanner and estimates of the low-temperature value of $\lambda$ for films with a $T_c$ of $\sim \SI{900}{\milli\kelvin}$ (See Appendix A for details). The absolute value of $\lambda$ is affected by uncertainty in the fit parameters, which are strongly correlated with $\lambda$, making a meaningful determination of the absolute value of $\lambda$ challenging (see Appendix A: Scanning SQUID Susceptometry for details). 

Fortunately, an important feature of Eq.~\ref{eq:eq1} is that its sole temperature dependence originates from $\lambda(T)$, which appears in a simple prefactor. This implies that by measuring $M$ as a function of temperature at a constant height above the sample, we can accurately extract relative changes in $\lambda(T)$, even in the presence of uncertainties in the geometric factors. To determine the temperature dependence of $\lambda$, we measure $M$ at a fixed position as a function of temperature, while maintaining light mechanical contact with the sample. Assuming that $M$ measured at \SI{50}{\milli \kelvin} reasonably approximates $\lambda_0 = \lambda(T=0)$ at a given position, we can obtain $\lambda(T)/\lambda_0 - 1 = \delta\lambda(T)\lambda_0$ from $\frac{M(z_o, T)}{M(z_o, T=0)} = \frac{\lambda_{0}^2}{\lambda^2(T)}$.

\begin{figure*}
    \centering
    \includegraphics[width=1.0\textwidth]{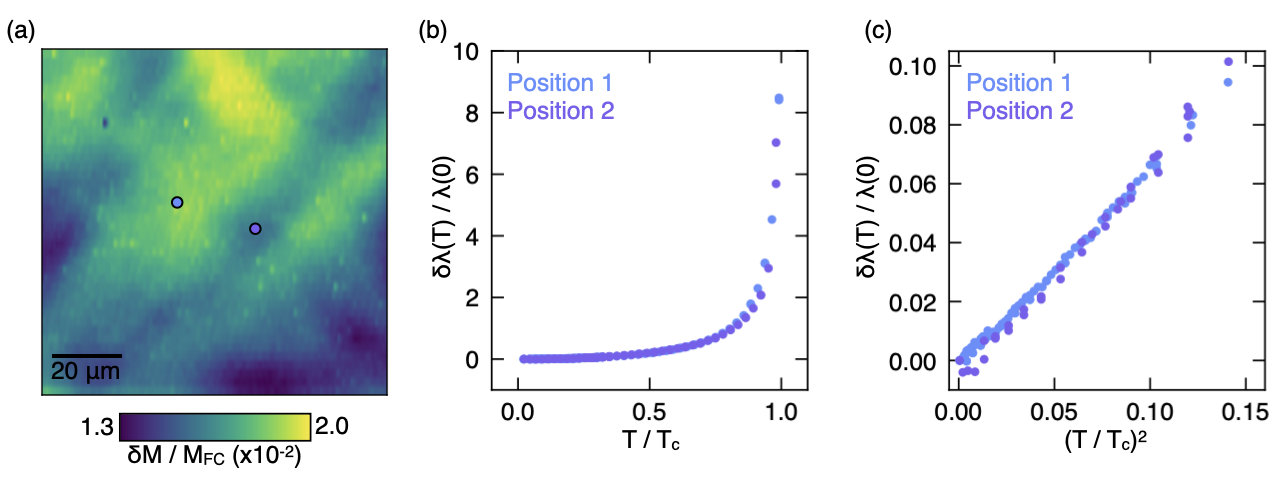}
    \caption{(a) Spatial map at \SI{20}{\milli\kelvin} of the local diamagnetic response. The two dots indicate the locations where detailed temperature dependent data were acquired. The left-most dot (blue) had a higher local superconducting transition temperature $T_c = \SI{0.90}{\kelvin}$ than the right dot (purple), with $T_c = \SI{0.85}{\kelvin}$. (b) $\delta \lambda \left(T\right) = \lambda(T) - \lambda(0)$, normalized by $\lambda(0)$ at the positions indicated in (a). (c) $\delta \lambda(T) / \lambda(0)$ plotted against $\left(T/T_c\right)^{2}$.
    }
    \label{fig:fig2}
\end{figure*}

Fig. 2a shows an image of the local magnetic response acquired by scanning several microns above the sample surface while maintaining the sample temperature at \SI{20}{\milli\kelvin}. The image reveals micrometer-scale variations in the local magnetic susceptiblity, and accordingly $\lambda$. Next, we conduct detailed temperature-dependent measurements of the penetration depth at two positions on the sample surface marked in Fig. 2a. In Fig. 2b, we show $\delta \lambda(T)/\lambda_0$ versus temperature for these two positions. We find a stronger diamagnetic response at base temperature and a higher local critical temperature, $T_c$ at Position 1 than at Position 2. For both positions, we observe no significant difference between data collected upon warming and cooling indicating that we sweep temperature sufficiently slow for the sample to thermalize.

In Fig. 2c, we plot $\delta \lambda(T)/\lambda_0$ as a function of $T^2/T_c^2$ for both positions. The data fall on a straight line, indicating that for $T< 0.4 T_c$, $\lambda(T)$ exhibits a $T^2$ temperature dependence. A temperature dependence of $\lambda(T) \sim T^2$ has been observed previously in high-quality single crystals \SRO{} using a tunnel diode oscillator technique \cite{bonalde2000temperature}, and more recently using scanning SQUID microscopy on single crystals under uniaxial strain \cite{mueller2023superconducting}. The power-law temperature dependence observed in our study and in the single crystal work provides evidence for nodes in the superconducting gap function of \SRO{}.

For nodal superconducting order parameters in the clean limit, a linear temperature dependence, $\delta \lambda(T) \sim T$ is expected. In \SRO{} single crystals, the $\delta\lambda(T) \approx T^2$ temperature dependence was attributed to non-local electrodynamics of the nodal quasiparticles \cite{kosztin1997nonlocal, bonalde2000temperature, mueller2023superconducting}. Impurity scattering can also change the temperature dependence of $\lambda(T)$ superconductors with nodes in the gap function. For example, for an order parameter with $d_{x^2-y^2}$ symmetry, strong impurity scattering is expected to give $\delta\lambda(T) \sim T^2$ \cite{hirschfeld1993effect}. In our thin-film samples, $T_c \approx \SI{1}{\kelvin}$ is suppressed compared to the highest $T_c \approx \SI{1.5}{\kelvin}$ observed in single crystals, and it is plausible that impurity scattering plays a larger role in determining the superconducting properties of the thin films than in bulk crystals. 

\begin{figure}
    \centering
    \includegraphics[width=0.5\textwidth]{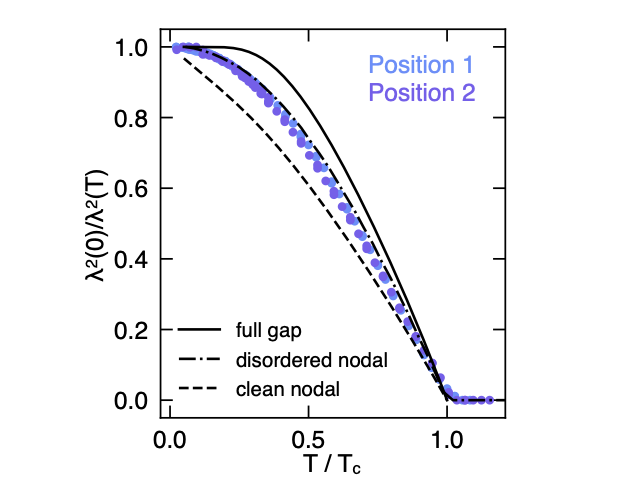}
    \caption{Normalized superfluid density $\lambda^{2}(0) / \lambda^{2}(T)$ measured at the two positions indicated in Fig. \ref{fig:fig2}a. Theoretical curves assuming a weak-coupling BCS superconductor with a single cylindrical Fermi surface are included for comparison.}
    \label{fig:fig3}
\end{figure}

To assess if scattering can explain the temperature dependence of the penetration depth, we examine the superfluid density across the full temperature range. The normalized superfluid density is directly related to the penetration depth through ${\rho_{s}(T)}/{\rho_0} = {\lambda_0^2}/{\lambda^2(T)}$, where $\rho_0$ is its zero temperature value. Figure \ref{fig:fig3} shows the normalized superfluid density for both positions measured in Figure \ref{fig:fig2}. Although we found local variations in both the magnitude of $\rho_{s}$ and $T_c$, we find that the normalized superfluid density collapses onto a single curve for both positions. Additional measurements collected on other areas of the sample, show the same temperature dependence of the normalized superfluid density (See Supplementary Information for Details).

In general, the temperature dependence of $\rho_s$ depends among other factors on the gap structure of the superconductor, the underlying band structure and scattering in the sample. Nevertheless, it is useful to compare the temperature dependence of $\rho_s$ to expectations from different simplified phenomenological models. In Fig. \ref{fig:fig3} we directly compare our measurements to calculations of the superfluid density for different superconducting gap functions assuming a single circular Fermi surface in the weak-coupling BCS limit (see Appendix: Calculation of the Superfluid Density for details). We find that neither the fully gapped model nor the d-wave models without scattering closely resemble our data.

% Range in T/Tc where T^2 is observed (compare to bulk crystals).
% given the level of disorder observed.

To include the effect of scattering on the superfluid density, we estimate the strength of pair-breaking scattering in our samples using the theory of Abrikosov and Gor'kov \cite{abrikosov1960contribution}, which has been used previously to successfully describe the dependence of $T_c$ on disorder in \SRO{} in both single crystals and epitaxial thin films \cite{mackenzie1998extremely}. Within this theory, $T_c$ satisfies
\begin{equation}
    \ln\left( \frac{T_{c0}}{T_c}\right) = 
    \Psi\left( \frac{1}{2} + \frac{\Gamma_{N}}{2 \pi T_c}\right) - \Psi\left(\frac{1}{2} \right),
\end{equation}
where $\Psi$ is the digamma function, $\Gamma_{N} = \hbar / 2 \tau k_{B}$ characterizes the strength of pair-breaking scattering with $\tau$ the corresponding scattering time, $k_{B}$ the Boltzmann constant, $\hbar$ the reduced Planck's constant, and $T_{c0}$ is the zero-disorder limit of $T_c$. Here we take $T_{c0} = \SI{1.5}{\kelvin}$, the highest value of $T_c$ observed in bulk crystals \cite{akima1999intrinsic}. In the thin film grown on LSAT studied here, we observe a suppressed $T_c \approx \SI{0.9}{\kelvin}$, corresponding to $\Gamma_{N} \approx \SI{0.7}{\kelvin}$. Here we assume that the Abrikosov-Gor'kov theory is applicable to both thin-film and bulk samples \cite{ruf2024controllable}, and that both types of sample have the same maximum $T_{c0}$. We note that \SRO{} thin films host defects that are not present in bulk samples. For example, step edges in the substrate due to a slight misorientation between the substrate surface and the (001) LSAT axis can cause out-of-phase boundaries that extend through a significant fraction of the film thickness \cite{nair2018demystifying, fang2021oscillations}.

% kB ~8.6e-5 eV/K ; hbar ~4.1e-15 eVs
Using the estimated value for $\Gamma_{N}$, we calculate the normalized superfluid density assuming a d-wave gap function with vertical line nodes, a single circular Fermi surface and including the effects of impurity scattering by following the approach of \cite{lee2017disorder} (details are provided in the Appendix). The result is included as line labeled ``disordered nodal" in Fig. 3, and provides much better agreement with our experimental data. We emphasize that this curve is not a fit to the experimental data, but a model for a disordered d-wave superconductor with the scattering rate estimated by the experimentally measured $T_c$. 

The agreement between this model and the data is remarkable given the simplicity of the Fermi surface and gap structure used in the calculations. Further refinement of this model by introducing multiband effects, the experimentally determined Fermi surface and a realistic pairing potential may improve the agreement between the model and data further. 

% This either over or under estimates the strength of the scattering... Probably under-estimates it

% Introduction to ring sample
We next discuss characterization of micrometer-scale rings fabricated from a \SRO{} thin film. Lithographically defined devices are enabled by thin film samples of \SRO{} and promising for future device-based tests of the superconducting order parameter symmetry.  We patterned an array of rings on a second \SRO{} thin film sample grown on an LSAT substrate using standard photolithography and argon ion milling. We present characterization of rings with an inner diameter of \SI{4}{\micro\meter} and an outer diameter of \SI{10}{\micro\meter}. 

\begin{figure}
    \centering
    \includegraphics[width=0.5\textwidth]{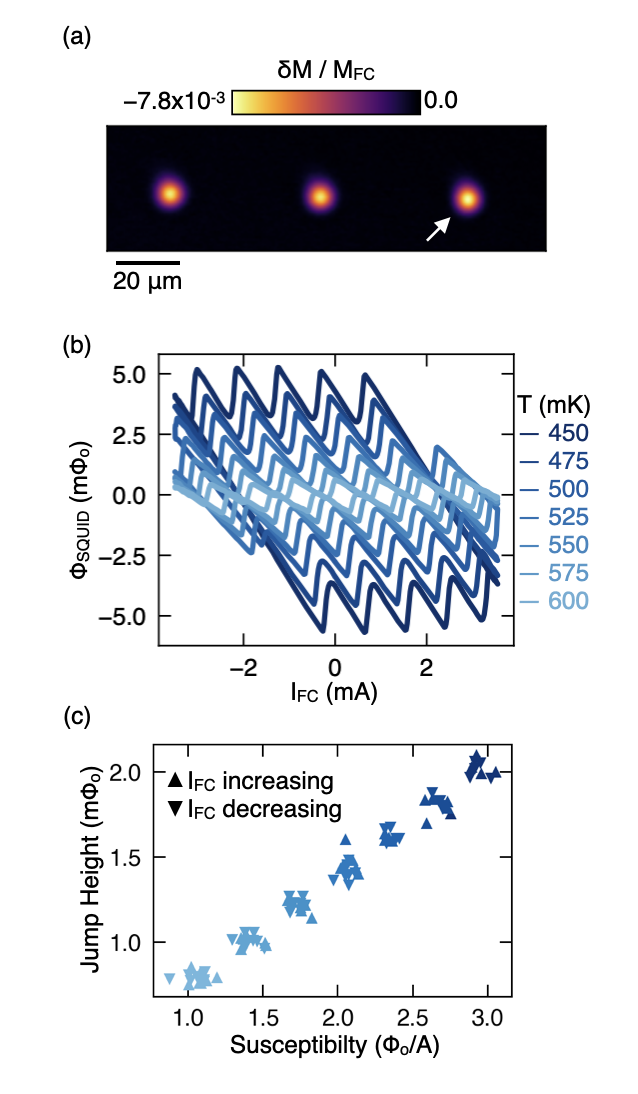}
    \caption{(a) Magnetic susceptibility image of three superconducting \SRO{} rings. (b) SQUID signal as a function of field coil current acquired at several temperatures below $T_c$ with the susceptometer positioned above the right-most ring in (a). The response of the ring is hysteretic with sharp jumps indicating of a discrete change in the ring fluxoid number superimposed on a diamagnetic background. (c) Size of the fluxoid jumps plotted against the diamagnetic susceptibility extracted from (b). Symbols indicate the direction of the field coil ramp for each jump. Color scale is the same as (b). As the temperature decreases, both the fluxoid jump size and the strength of the ring diamagnetic response increase.}
    \label{fig:fig4}
\end{figure}

In Fig.~\ref{fig:fig4}a we show a magnetic susceptibility image of three representative \SRO{} rings. To characterize the magnetic response of the rings in detail, we position the susceptometer above the ring and apply a current oscillating sinusoidally at a few Hz to the field coil while monitoring the ring response with the SQUID. We average the signal from up to a few hundred oscillations of the field coil current in order to achieve a high signal-to-noise ratio. A background signal, measured with the SQUID retracted from the sample at each temperature, is subtracted from the data. Fig. 4b shows the magnetic response of the right-most ring in Fig.~\ref{fig:fig4}a at several temperatures below $T_c$. As the field coil current is swept, we observe a linear response which is periodically interrupted by abrupt jumps in the SQUID signal. At each temperature measured, the response of the ring to the field coil current is hysteretic. The size of the hysteresis loop, the slope of the ring response, and the size of the jumps in the signal all decrease as the temperature of the sample is raised towards $T_c$.

% Phenomenology of the ring signals
To understand the magnetic response of the rings, we turn to the following London equation which determines the current distribution in the ring,
\begin{equation*}
    \vec{j} = -\frac{1}{\mu_0 \lambda^2} \left[\frac{\Phi_0}{2\pi}\left( \nabla \theta\right) + \vec{A} \right],
\end{equation*}
where $\vec{j}$ is the supercurrent density, $\mu_0$ is the vacuum permeability, $\Phi_0 = \frac{h}{2e}$ is the superconducting flux quantum, $\theta$ is the order parameter phase and $\vec{A}$ is the magnetic vector potential. Integrating over the film thickness and introducing cylindrical coordinates $r$ and $\phi$ with respect to the center of the ring gives,
\begin{equation}
    g_{\phi}(r) = \frac{d}{\mu_o \lambda^2} \left[ \frac{\Phi_o}{2\pi}\frac{N}{r} - A_{\phi}(r)\right],
    \label{eq:sc_current}
\end{equation}
where we have enforced single-valuedness of the superconducting order parameter with the integer $N$ the number of vortices trapped inside the ring, and $g_{\phi}(r)$ the sheet current density flowing along the azimuth $\phi$ of the ring. $A_{\phi}$ is the azimuthal component of the magnetic vector potential.

This model captures the main qualitative features of our data. We interpret the sudden jumps in Fig.~\ref{fig:fig4}b as changes in the vortex number $N$ via the transport of a superconducting vortex accross the wall of the ring. Changes in $N$ generate discontinuous changes in the current flowing through the ring, which produce jumps in the flux coupled into the SQUID pickup loop. Away from these transitions, the flux threading the ring changes linearly with the field coil current corresponding to a linear change in $A_{\phi}$ which increases the diamagnetic screening current in the ring and consequently the flux coupled into the SQUID pickup loop. 

A quantitative prediction of the jump height requires a self-consistent solution of $g_{\rho}(r)$ in the presence of the $A_{\phi}$ generated by the field coil excitaion. For our samples and susceptometer, where the field coil dimensions are comparable to the ring dimensions, these calculations must be carried out numerically \cite{kirtley2003fluxoid}. Eq.~\ref{eq:sc_current} shows that both the size of the jumps when $N$ changes by one and the strength of the diamagnetic screening at a given temperature $T$ are proportional to $1/\lambda^2$. In Fig.~\ref{fig:fig4}c we show the size of the jumps at each $T$ plotted against the slope of the response away from the jumps, which we attribute to diamagnetic screening. Consistent with Eq.~\ref{eq:sc_current}, we find that the jump height and the strength of diamagnetic screening increase monotonically and proportional to each other as the temperature is lowered and $1/\lambda^2$ increases. We find that the number of vortex transitions goes down as the temperature is lowered, consistent with vortex motion accross the ring wall. As the superfluid density in the ring increases, the energetic barrier for a vortex tunneling accross the ring increases. The field coil current required to induce a transition therefore increases and the hysteresis in the ring signal becomes more pronounced at lower temperatures. We investigate the temperature range between \SI{0.650}{\kelvin} and \SI{0.550}{\kelvin}, near the superconducting transition of the rings $T_{c} = \SI{0.725}{\kelvin}$, where fluxoid transitions are readily accessed with field coil currents of a few mA.

% Temperature dynamics?

% Conclusion -- a summary of the findings

% Next -- devices and biaxial tension
In summary, we find that the penetration depth in thin-film \SRO{} shows a $\lambda(T) \propto T^2$ dependence at low temperatures similar to reports in single crystal bulk samples. This suggests, as is expected, that the gap structure in our thin films is comparable to single crystal samples and features nodes. Analysis of the full temperature dependence of the superfluid density shows that it can be explained by a nodal superconducting gap combined with scattering in our samples. While this does not exclude non-local effects in the Meissner screening, scattering is in our case the simplest explanation. In addition, we provide the first magnetic characterization of mesoscopic superconducting rings lithographically patterned from a \SRO{} thin-film. Looking forward, our work shows that the delicate superconducting state of \SRO{} can be preserved in high-quality thin film devices, which is an important first step towards device-based tests of the order parameter symmetry. 
Furthermore, the effect of biaxial strain on the superconducting state in \SRO{} may be studied by choosng suitable substrates. Uniaxial strain has proven to be an effective tuning parameter for superconductivity in bulk \SRO{} \cite{hicks2014strong, steppke2017strong}. Biaxially strained thin-film samples \cite{burganov2016strain} may exhibit a qualitatively different ground state than their uniaxially strained counterparts \cite{liu2018superconductivity}. The experimental approach demonstrated in this work will provide direct access to new superconducting or magnetic states that emerge in such samples.

\begin{acknowledgments}
Scanning SQUID measurements were primarily supported through the Cornell University Materials Research Science and Engineering Center DMR-1719875. Thin film growth was supported by the National Science Foundation Platform for the Accelerated Realization, Analysis, and Discovery of Interface Materials (PARADIM) under Cooperative Agreement No. DMR-2039380. In addition, this research was funded in part by the Gordon and Betty Moore Foundation’s EPiQS Initiative through Grant Nos. GBMF3850 and GBMF9073 to Cornell University. Sample preparation was facilitated in part by the Cornell NanoScale Facility, a member of the National Nanotechnology Coordinated Infrastructure (NNCI), which is supported by the National Science Foundation (Grant No. NNCI-2025233). K.M.S. acknowledges the support of Air Force Office of Scientific Research Grant No. FA9550-21-1-0168 and National Science Foundation DMR-2104427.
\end{acknowledgments}

\appendix
\section{Scanning SQUID Susecptometry}
Our SQUID susceptometer features a counter-wound field coil pickup loop pair that allows us to null the SQUID response to the field coil excitation. To perform local magnetic susceptibility measurements, the SQUID is retracted as far as possible from the sample surface (about \SI{30}{\micro\meter}) and the susceptometer is balanced by adjusting the current flowing in the two arms of the field coil until the flux coupled into the SQUID pickup loops is minimized. Any remaining SQUID signal after balancing is recorded as an offset signal which represents the response of the SQUID in to the field coil in the absence of the sample. This offset is then subtracted from the height sweep data used to characterize the susceptometer geometry in Fig. 1b and from all height sweep data used to extract the penetration depth and superfluid density (Figs. 2-3). No offset was subtracted from imaging measurements of the local susceptibility (Fig 2a, Fig 4a) as these images are used to characterize the spatial variations in the diamagentic response of the sample.

To avoid spurious signals from vortex motion during the penetration depth measurements, the susceptibility data were collected conducted under near zero field conditions. A superconducting solenoid was used to adjust the out-of-plane magnetic field  threading the sample until no vortices were observed within the $\SI{140} \times \SI{140}{\micro\meter}^2$ field of view after warming and cooling the sample and SQUID through their respective critical temperatures. This procedure places a rough upper bound of $\frac{\Phi_{0}}{\SI{140} \times \SI{140}{\micro\meter}^2} \approx \SI{100}{\nano\tesla}$ on the magnetic flux density in the sample environment. Before and after conducting the temperature sweeps used to acquire the magnetic susceptibility data, the SQUID was used to acquire an image confirming that vortices were absent from the measurement area.

In the main text, we utilized an analytical expression (Eq. \ref{eq:eq1}) relating the magnetic penetration depth to the geometry of our magnetic susceptibility measurement. Although the data is well-fit by this model, it is strictly only valid deep in the Pearl limit where, $\lambda >> d$. Muon spin relaxation measurements of the zero-temperature penetration depth in \SRO{} have found that at the lowest temperatures, $\lambda(T) \approx \SI{190}{\nano\meter}$ \cite{riseman1998observation}, which is only a few times our film thickness of \SI{35}{\nano\meter}. In the thin-film samples that we measure here, $T_c$ is substantially suppressed from the values observed in the best single crystals. Using calculations of the zero temperature superfluid density in the presence of disorder, described below, we estimate that $\lambda(0)$ in our samples ($T_c \approx \SI{900}{\milli\kelvin}$, $\Gamma_N \approx \SI{0.7}{\kelvin}$) is a factor of $\sim 1.7$ larger than $\lambda(0)$ in the best single crystals. We therefore estimate that at all temperatures measured, $\lambda(T) \gtrsim 10d$.

To extract the temperature dependence of $\lambda(T)$ from the magnetic susceptibility measurements, we observe that the expression Eq. \ref{eq:eq1} may be separated into the product of a temperature-dependent part and a temperature-independent part that depends only on the geometric configuration of the SQUID and sample,
\begin{equation}
    \frac{M(z, T)}{M_{FC}} 
    = -\frac{a d}{2 \lambda^2(T)} \left[ 1 - \frac{2 z}{\sqrt{a^2 + 4z^2}}\right]
    = \frac{A_{geo}(z)}{\lambda^2(T)}.
\end{equation}
Where $A_{geo}(z) = -\frac{ad}{2}\left[1 - \frac{2z}{\sqrt{a^2 + 4z^2}} \right]$, encodes the sample-susceptometer geometry. From this expression, the temperature dependence of the penetration depth, $\delta \lambda(T) / \lambda_o$ may be written,
\begin{equation}
    \frac{\delta\lambda(T)}{\lambda_o} 
    = \frac{\lambda(T) - \lambda_o}{\lambda_o} 
    = \sqrt{\frac{M(z_o, T=0)}{M(z_o, T)}} - 1,
\end{equation}
allowing $\delta \lambda(T) / \lambda_o$ to be directly calculated from measurements of $M(z_o, T)$.

\section{Superfluid Density Calculations}
In this work we calculate the superfluid density in the weak-coupling limit of BCS theory. 

To calculate the superfluid density in the presence of disorder, we turn to the self-consistent t-matrix approximation \cite{hirschfeld1993effect, lee2017disorder}. Within the approximation, impurities are treated as isotropic point scatters. For simplicity, we again model the system as a single circular Fermi surface, and choose the simplest separable pairing potential $V_{\textbf{k}, \textbf{k'}}$ with a d-wave form factor,
\begin{equation}
    V_{\textbf{k}, \textbf{k'}} = V_{0} \Omega_{\textbf{k}} \Omega_{\textbf{k'}}.
\end{equation}
Where,
\begin{equation}
    \Omega_{\textbf{k}} \propto \cos{k_x a} - \cos{k_y a},
\end{equation}
with $a$ the lattice spacing. Given this Fermi surface and pairing potential, we find solutions to the gap equation,
\begin{equation}
    \Delta_{\textbf{k}} = 2 \pi T \sum_{\omega_n > 0}^{\omega_0}
    \left< V_{\textbf{k}\textbf{k}'} 
    \frac{\Delta_{\textbf{k}'}}{\left( \tilde{\omega}_n^2 + \Delta_{\textbf{k}'}^2 \right)}
    \right>_{FS}.
\end{equation}
Here, $\Delta_{\textbf{k}} = \psi(T) \Omega_{\textbf{k}}$ is the superconducting gap, with $\psi(T)$ the temperature-dependent gap amplitude. $\omega_n = 2 \pi T \left(n + \frac{1}{2} \right)$ are the fermionic Matsubara frequencies and $\left< \cdot \cdot \cdot \right>_{FS}$ denotes an average over the Fermi surface. Adding disorder to the system renormalizes the Matsubara frequencies:
\begin{equation}
    \tilde{\omega}_n = \omega_n + \pi \Gamma 
    \frac{
    \left< N_{\textbf{k}}(\tilde{\omega}_n) \right>_{FS}
    }
    {
    c^2 + \left< N_{\textbf{k}}(\tilde{\omega}_n) \right>_{FS}
    }.
    \label{eq:w_rn}
\end{equation}
Where $c$ is the cotangent of the scattering phase shift and $\Gamma$ parameterizes the density of scattering sites. The limits $c<<1$ and $c>>1$ correspond to strong (unitary) and weak (Born) scattering respectively. For this work, we observe that impurities appear to act as strong pair-breaking scatterers in \SRO{} \cite{mackenzie1998extremely, ruf2024controllable} and take $c=0$. Within weak-coupling BCS theory, the gap equation may be solved by finding the $\tilde{\omega}_n$ and $\psi(T)$ that satisfy,
\begin{equation}
    \ln{\left(\frac{T_{c0}}{T}\right)} = 2 \pi T \sum_{\omega_n>0}^{\infty}\left(\frac{1}{\omega_n} -
    \left< 
    \frac{\Omega_{k}^2}{\left( \tilde{\omega}_{n}^2 + \psi^2 \Omega_k^2 \right)^{1/2}}
    \right>_{FS}
    \right).
    \label{eq:self_const}
\end{equation}

Equations \ref{eq:self_const} and \ref{eq:w_rn} may be solved self-consistently at a range of temperatures to obtain $\tilde{\omega}_n$ and $\psi(T)$. Once $\tilde{\omega}_n$ and $\psi(T)$ are known, the superfluid density in the presence of disorder may be calculated,

\begin{equation}
    \frac{\rho_{s}(T)}{\rho_{s00}} = 2\pi T \sum_{\omega_n > 0}^{\infty}
    \left< \left< 
    \frac
    {\Delta_{\textbf{k}}^2}
    {\left(\tilde{\omega}_{n}^2 + \Delta_{\textbf{k}}^{2} \right)^{3/2}}
    \right> \right>_{FS}.
\end{equation}

$\left< \left< \cdot \cdot \cdot \right> \right>_{FS}$ denotes a velocity-weighted average over the Fermi surface, and in the case of a circular Fermi surface,
\begin{equation}
    \left<\left< A(\phi)\right>\right>_{FS} = 
    \frac
    {\int_{0}^{2\pi} A(\phi) v_{F,x}^2}
    {\int_{0}^{2\pi} v_{F,x}^2}.
\end{equation}
Examples of calculated $\psi(T)$ and $\rho_s(T)$ under different impurity scattering strengths $\Gamma_N$ are presented in the Supplementary Information.

\section{Sample Growth}

The \SRO{} thin film was grown in a Veeco Gen10 molecular-beam epitaxy (MBE) system on a $(\mathrm{LaAlO_3})_\mathrm{0.29}-(\mathrm{SrAl}_{0.5}\mathrm{Ta}_{0.5}\mathrm{O}_3)_{0.71}$ substrate from CrysTec GmbH. The substrate used for the growth was screened to have a miscut of less than \SI{0.05}{\degree}, which is important to reduce the formation of out-of-phase boundaries. The films were grown at a substrate temperate of \SI{810}{\degreeCelsius} as measured using an optical pyrometer operating at \SI{1550}{\nano\meter}. Elemental strontium (99.99\% purity) and elemental ruthenium (99.99\% purity) evaporated from a low-temperature effusion cell and a Telemark electron beam evaporator, respectively, were used for growing the \SRO{} film. The films were grown with a strontium flux of \SI{2.6e13}{\atoms\per\centi\meter\tothe{2}\per\second} and a ruthenium flux of \SI{1.8e13}{\atoms\per\centi\meter\tothe{2}\per\second} in a background of distilled ozone ($\sim$80\% $\mathrm{O_{3}}$ + 20\% $\mathrm{O_{2}}$ made from oxygen gas with 99.994\% purity.) The background oxidant pressure during growth was \SI{3e-6}{\torr}. At the end of the growth the strontium and ruthenium shutters were closed simultaneously, and the sample was cooled down to below \SI{250}{\degreeCelsius} in a background pressure of distilled ozone of \SI{1e-6}{\torr}. Further details of the adsorption-controlled growth conditions for the growth of \SRO{} thin films by MBE can be found elsewhere \cite{nair2018demystifying}.

\section{Fabrication Procedure}

The \SRO{} films grown by MBE were subsequently patterned into rigs for scanning SQUID microscopy using standard photolithography, sputter deposition, and ion milling techniques. First, a Pt meander structure used for navigation on the sample was defined using photolithography. Next, \SI{25}{\nano\meter} of platinum, with \SI{5}{\nano\meter} titanium adhesion layer, was sputtered onto the \SRO{} film with an AJA sputtering tool, followed by a standard lift-off processes. A second photolithography step was used to define the ring  geometry, followed by ion milling with an AJA ion mill to remove the excess \SRO{} film.

\nocite{*}

\bibliography{references}% Produces the bibliography via BibTeX.

\end{document}

\renewcommand{\thefigure}{S\arabic{figure}}

\preprint{APS/123-QED}

\title{Supplementary Information for Local Magnetic Measurements of \SRO{} Thin Films}

\maketitle

\section{Additional measurements of the superfluid density}

\begin{figure*}
    \centering
    \includegraphics[width=1.0\textwidth]{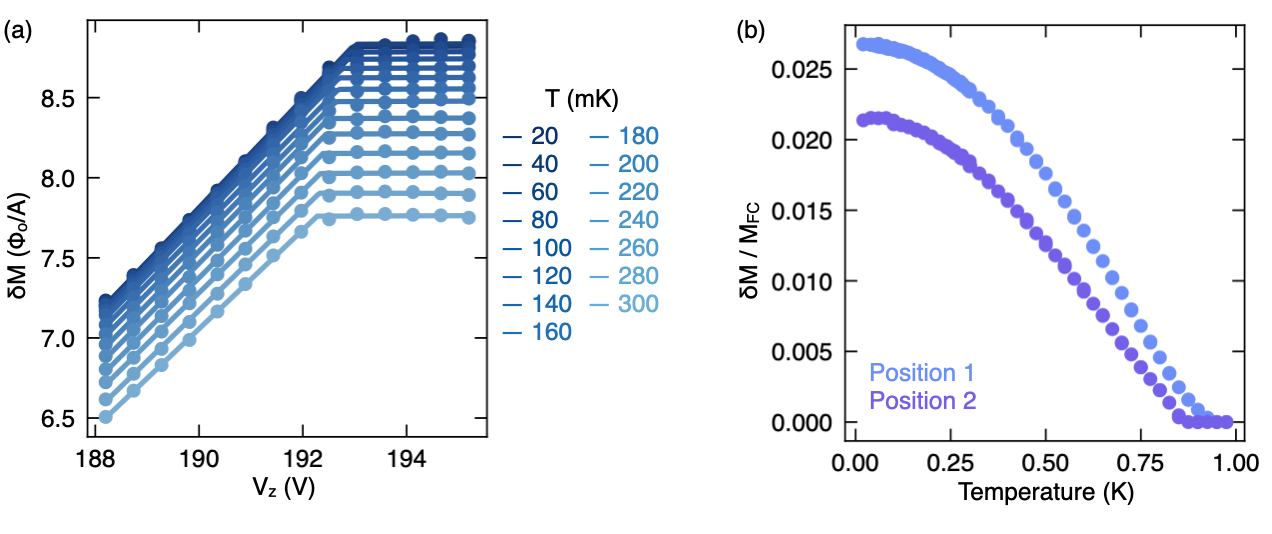}
    \caption{(a) SQUID-field coil mutual inductance data collected at Position 1 indicated in Figure 2(a) in the main text. As the SQUID is swept towards the sample surface, the strength of the signal detected by the SQUID increases until the SQUID makes mechanical contact with the sample. A piece-wise linear fit to the data is used to extract the the $V_z$ where the SQUID makes contact as well as $M(z_o, T)$, the value of the SQUID-field coil mutual inductance when the SQUID is in contact with the sample. As the sample temperature increases, the diamagnetic screening from the sample becomes weaker. (b) Temperature dependence of $M(z_o, T)$ at the positions indicated in Figure 2(a) in the main text. $M(z_o, T)$ is extracted from data acquired by the procedure used in (a) over a wider range of temperatures.
    }
    \label{fig:si_1}
\end{figure*}

In Fig S1, we provide raw magnetic susceptibility data acquired with the scanning SQUID, used to generate the curves in Fig 2 of the main text.

To check the consistency of our local measurements of the superfluid density, we compared the results presented in the main text to additional measurements performed in a new field of view on the same sample (Fig S2). The new field of view was separated by approximately \SI{1}{\milli\meter} from the field of view presented in the main text.

In total, we have performed detailed temperature sweeps at five different locations an the sample surface. We collected these sweeps at locations where the local $T_c$ and $n_s$ where higher than average as well as locations where the local $T_c$ and $n_s$ were lower than surrounding points. Regardless of where we collected our temperature dependent data, we found that the normalized superfluid density to collapse onto a single curve (Fig S2e).

\begin{figure*}
    \centering
    \includegraphics[width=1.0\textwidth]{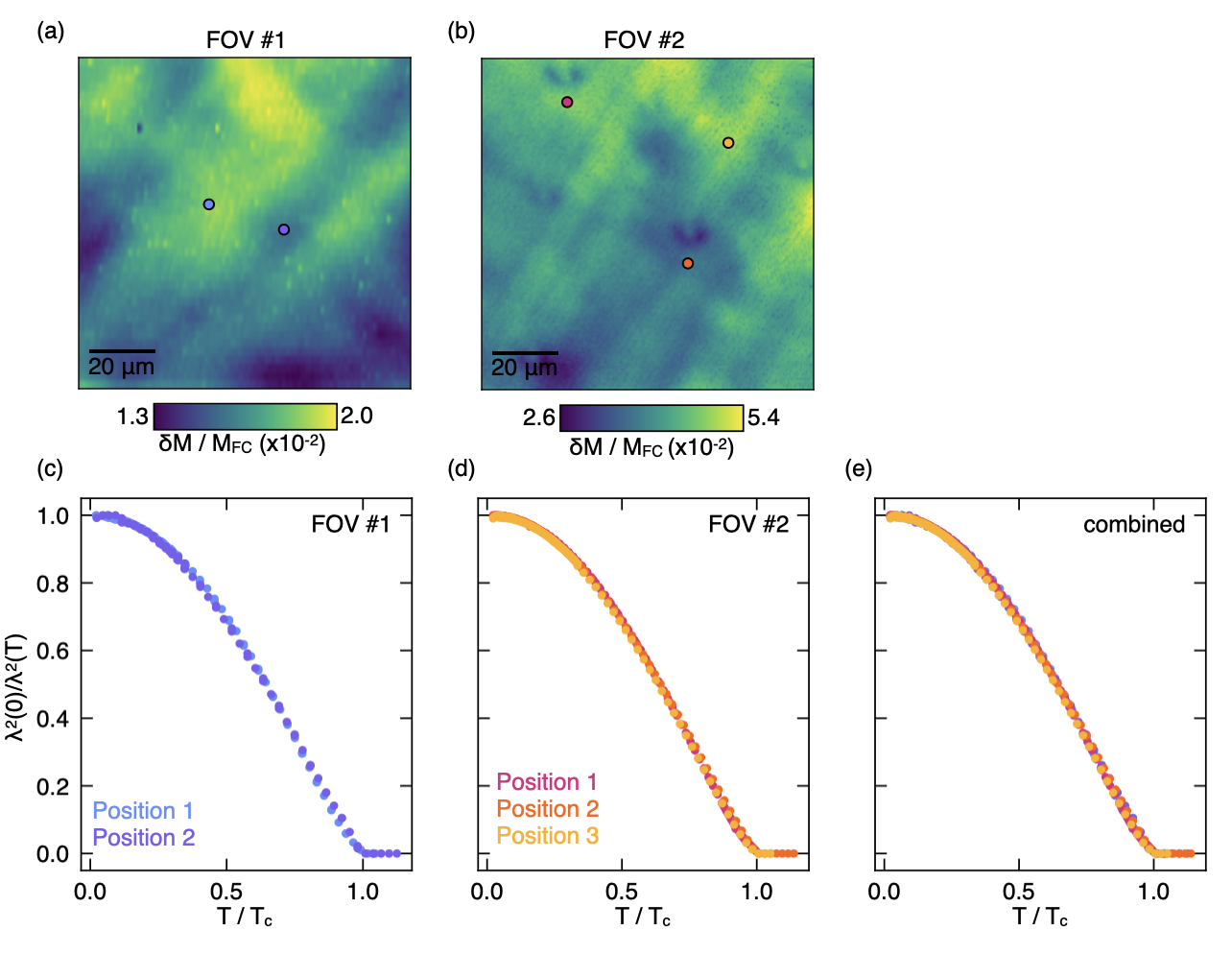}
    \caption{(a, b) Magnetic susceptibility images of both fields of view where detailed temperature-dependent data were collected. (a) is reproduced from Fig 2. (c) Normalized superfluid density reproduced from Fig 3. (d) Same as (c) except data was collected in FOV 2 at the points indicated in (b). (e) Direct comparison between the superfluid density measured at all five positions.
    }
    \label{fig:si_1}
\end{figure*}

\section{Fluxoid transitions in additional rings}
We provide an optical image of the lithographically patterned \SRO{} ring sample in Fig. S3. We have performed detailed characterization of rings with two different lithographic dimensions. For each ring size, we have measured several different rings. In Fig S4, we include fluxoid transition data for additional rings of the same dimensions as those presented in the main text. Although we identified many rings that exhibited qualitatively similar fluxoid transitions, the detailed temperature dependent behavior was found to vary between rings.

In Fig S5, we include measurements of fluxoid transitions in several rings of smaller dimensions than those presented in the main text.

\section{Sensitivity of the superfluid density to $\Gamma_{N}$}
In the main text, we used the local $T_c$ measured with the scanning SQUID to estimate the value of the impurity scattering rate $\Gamma_N$ in our samples. Using this value of $\Gamma_N$ we fould good agreement between a simple model for the superfluid density which includes both nodes in the gap fundction and disorder. In Fig S6, we show that the results of this calculation are not particularly sensitive to the value of $\Gamma_N$ that we choose. In particular, for values of $\Gamma_N$ appropriate for our samples, changes in $\Gamma_N$ result in large changes in the absolute value of the superfluid density at low temperatures, but do not substantially change the shape of the normalized superfluid density which we plot in Fig 3 of the main text.

\begin{figure*}
    \centering
    \includegraphics[width=0.5\textwidth]{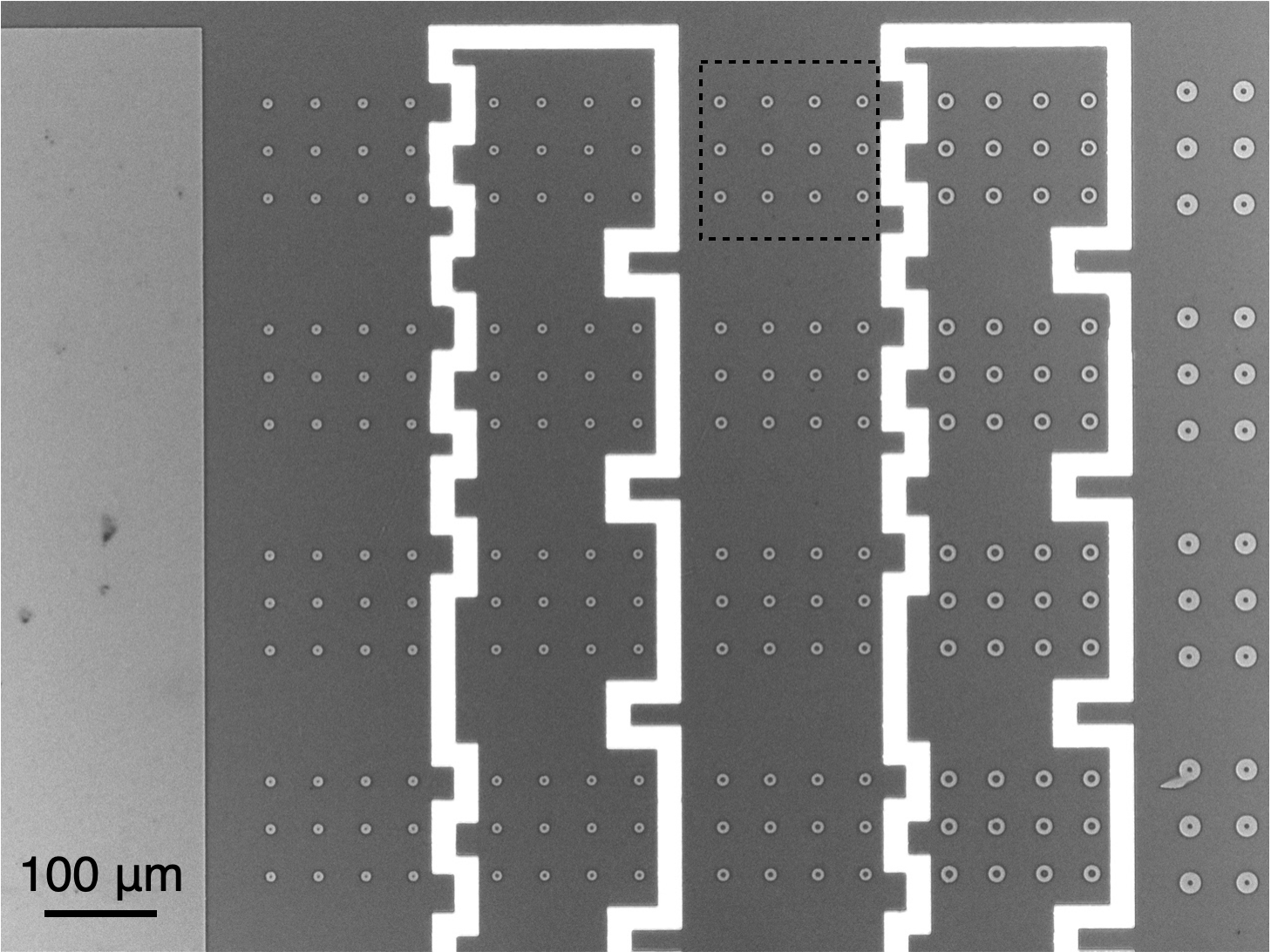}
    \caption{Optical microscope image of the lithographically defined array of \SRO{} rings. Measurements on rings located in the dashed box (inner diameter \SI{4}{\micro\meter} and outer diameter \SI{10}{\micro\meter} respectively) are shown in Fig. 4 and Fig. S4
    }
    \label{fig:si_1}
\end{figure*}

\begin{figure*}
    \centering
    \includegraphics[width=1.0\textwidth]{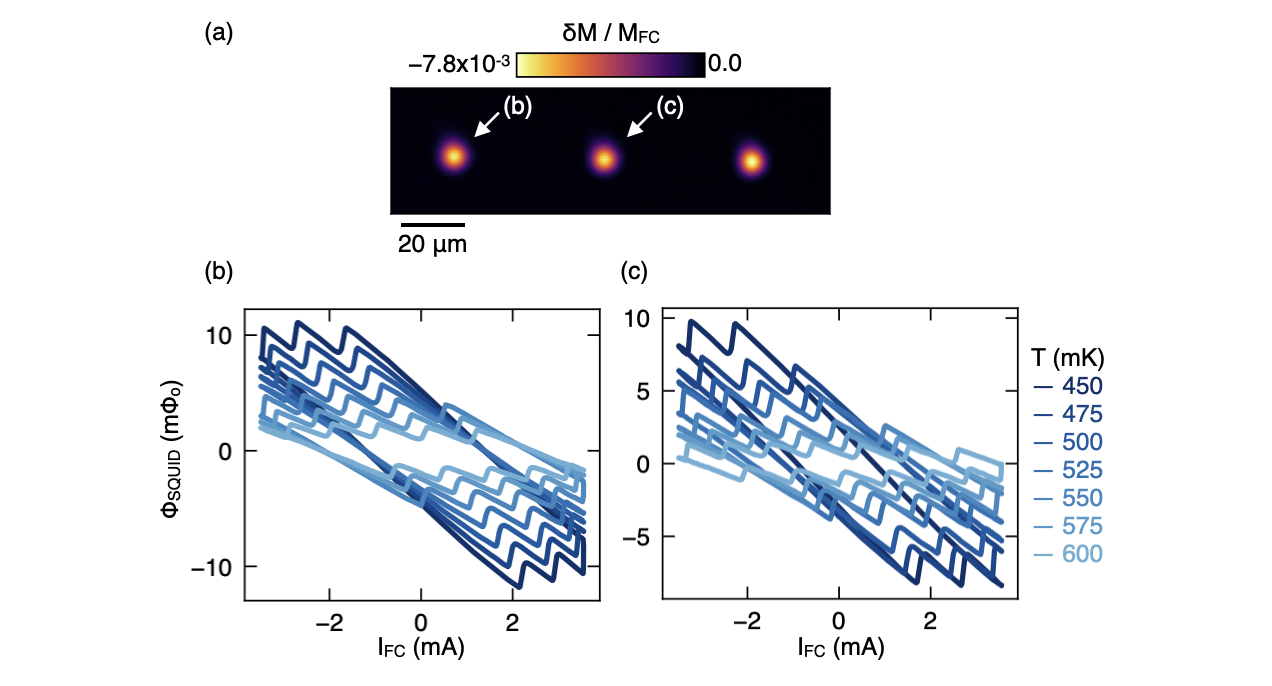}
    \caption{(a) Magnetic susceptiblity image for 3 \SRO{} rings. Reproduced from Fig 4 in the main text. (b) Temperature dependence of fluxoid transitions for the left-most ring indicated in (a). (c) Same as (b) but for the center ring in (a). Although both rings exhibit fluxoid transitions and temperature-dependent hysteresis, the details of their fluxoid transitions are different.
    }
    \label{fig:si_2}
\end{figure*}

\begin{figure*}
    \centering
    \includegraphics[width=1.0\textwidth]{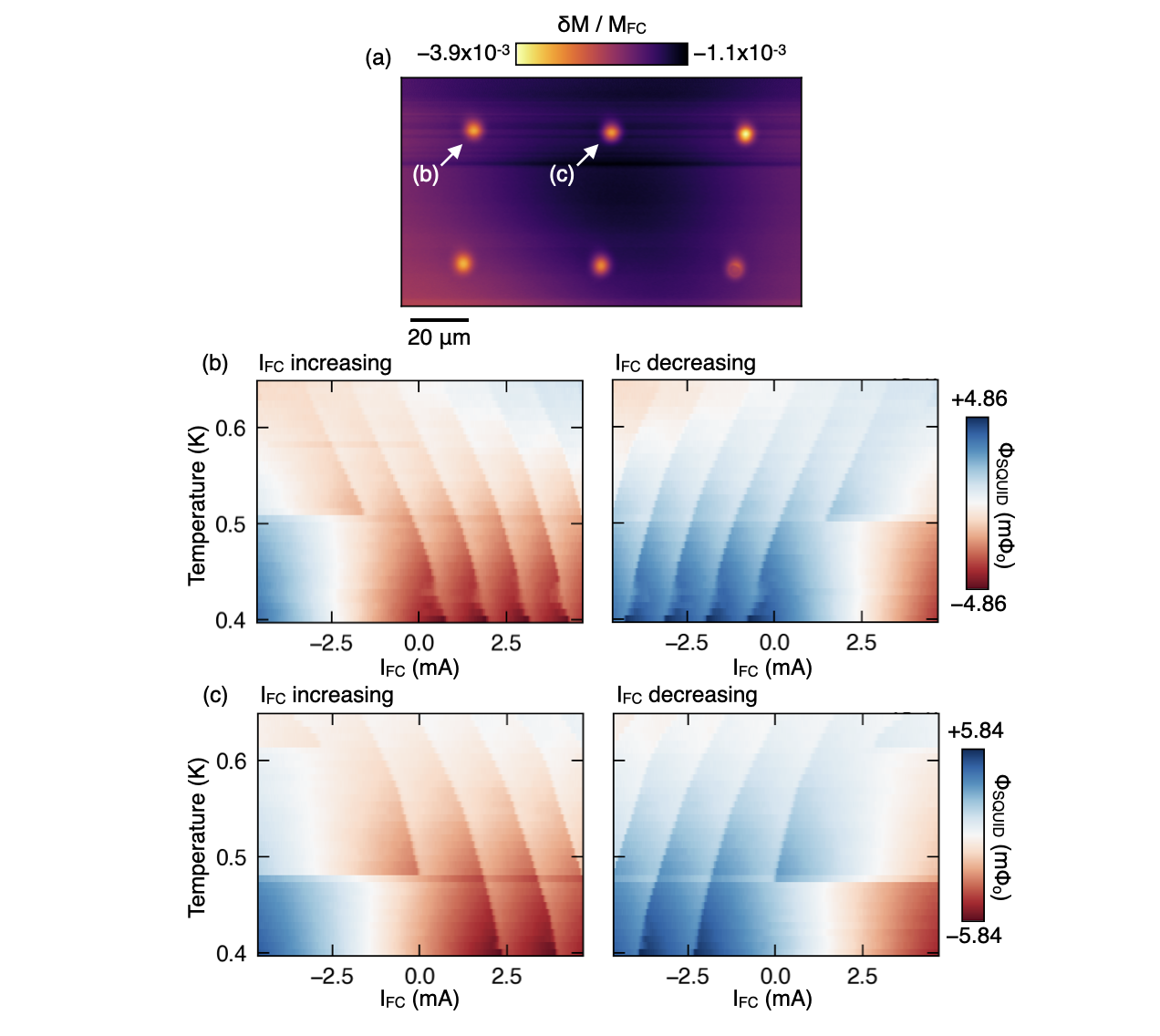}
    \caption{(a) Magnetic susceptibility image of six smaller \SRO{} rings with inner diameter \SI{2}{\micro\meter} and outer diameter of \SI{4}{\micro\meter}. (b) Temperature dependant fluxoid transitions for the top-left ring indicated in (a). For clarity, the SQUID signal for increasing $I_{FC}$ is plotted on the left and the SQUID signal for decreasing $I_{FC}$ is plotted on the right. (c) Same as (b) but for an additional ring indicated in (a). Although both rings exhibit qualitatively similar magnetic behavior, the details of the fluxoid transitions remain different.
    }
    \label{fig:si_3}
\end{figure*}

\begin{figure*}
    \centering
    \includegraphics[width=1.0\textwidth]{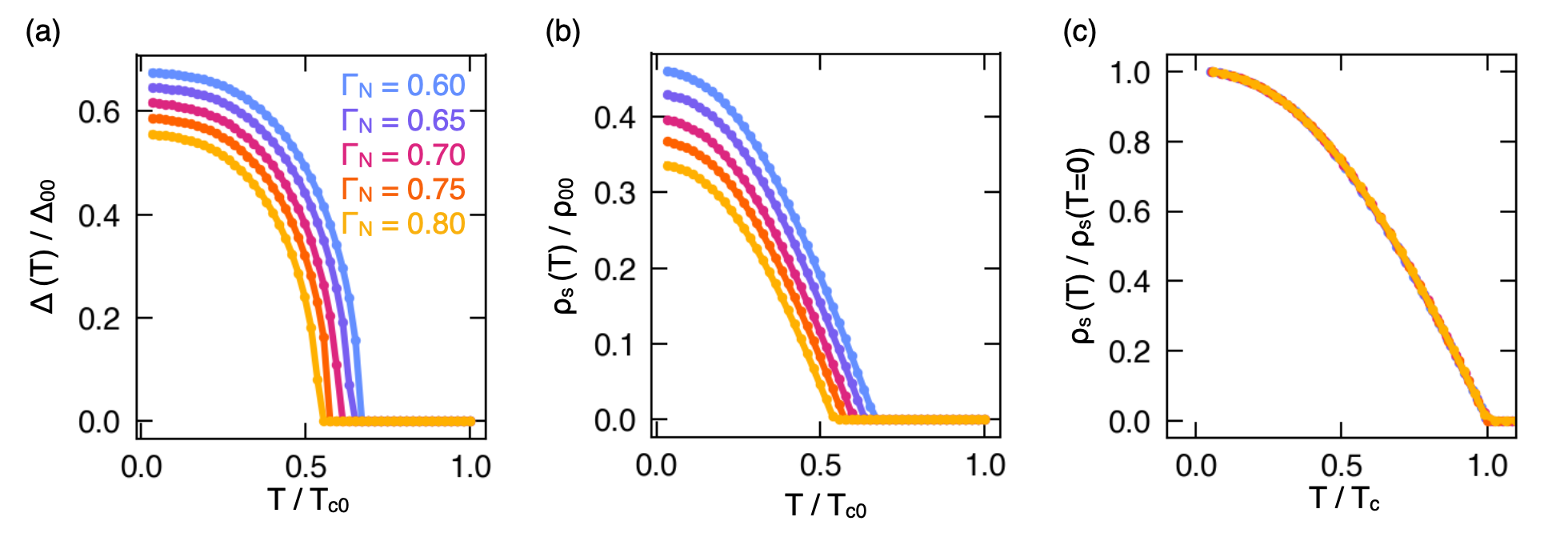}
    \caption{Sensitivity of the superfluid density calculations to the scattering rate $\Gamma_N$. (a) Self-consistent gap amplitude calculated as a function of $T$, normalized by the zero temperature, zero diorder gap amplitude $\Delta_{00}$. The calculation is repeated at four representative values of $\Gamma_N$ near the value estimated from the local $T_c$ of the thin film. (b) Temperature dependent superfluid density calculated using the gap amplitudes in (a), normalized by the zero temperature, zero disorder superfluid density $\rho_{00}$. (c) Normalized superfluid denstiy as a function of temperature calculated in the same way as Fig 3 in the main text. Each curve is normalized by its own zero temperature value and critical temperature. For the values of $\Gamma_N$ appropriate for our samples, the shape of the curve is insensitive to the exact value of $\Gamma_N$.
    }
    \label{fig:si_3}
\end{figure*}

\bibliography{references}% Produces the bibliography via BibTeX.